\documentclass[10pt, conference, compsocconf]{IEEEtran}
\usepackage{subfigure}

\usepackage[utf8]{inputenc}
\usepackage{amsmath,amssymb}

\usepackage{graphicx}
\usepackage{subfigure}
\usepackage{color}
\usepackage{array}

\usepackage{algorithm}
\usepackage{algorithmic}

\begin{document}
%
% --- Author Metadata here ---
% \conferenceinfo{SIMUTools 2009}{March 2-6, Rome, Italy}
%\setpagenumber{50}
% \CopyrightYear{2009} % Allows default copyright year (2002) to be over-ridden - IF NEED BE.
%\crdata{1-60558-090-6/08/07}  % Allows default copyright data (X-XXXXX-XX-X/XX/XX) to be over-ridden.
% --- End of Author Metadata ---

% \title{On Modeling the Dynamics of Self-Organizing Opportunistic Networks}
\title{Scale-Free Opportunistic Networks: is it Possible?}

\author{\IEEEauthorblockN{Stefano Ferretti, Vittorio Ghini}
\IEEEauthorblockA{Department of Computer Science, University of Bologna\\
Bologna, Italy\\
\{sferrett, ghini\}@cs.unibo.it}
}

% \numberofauthors{1}
% \author{
% \alignauthor Stefano Ferretti\\
%        \affaddr{Department of Computer Science, University of Bologna}\\
%        \affaddr{Mura Anteo Zamboni 7, 40127, Bologna, Italy}\\
%        \email{sferrett@cs.unibo.it}
% }

\date{}

\maketitle

\begin{abstract}
The coupling of scale-free networks with mobile unstructured networks is certainly unusual. In mobile networks, connections active at a given instant are constrained by the geographical distribution of mobile nodes, and by the limited signal strength of the wireless technology employed to build the ad-hoc overlay. This is in contrast with the presence of hubs, typical of scale-free nets. However, opportunistic (mobile) networks possess the distinctive feature to be delay tolerant; mobile nodes implement a store, carry and forward strategy that permits to disseminate data based on a multi-hop route, which is built in time, when nodes encounter other ones while moving. In this paper, we consider opportunistic networks as evolving graphs where links represent contacts among nodes arising during a (non-instantaneous) time interval. We discuss a strategy to control the way nodes manage contacts and build ``opportunistic overlays''. Based on such an approach, interesting overlays can be obtained, shaped following given desired topologies, such as scale-free ones.
\end{abstract}
%\nodate{}

\begin{IEEEkeywords}
Opportunistic Networks, Scale-Free Networks, Self-organization
\end{IEEEkeywords}

% \category{C.2.1}{Computer-Communication Networks}{Network Architecture and Design}%
% \terms{Algorithms, Performance, Design}
% \keywords{Opportunistic Networks, Scale-Free Networks, Self-organization} % NOT required for Proceedings

\section{Introduction}

Opportunistic networks are very dynamical systems, characterized by intermittent contacts among mobile nodes and frequent partitions \cite{boldrini,chaintreau}. The model of interaction differs from classic communication paradigms, usually based on a prolonged end-to-end connectivity \cite{gridpeer}. 
At a given instant, a complete path from a source to a destination might not exist \cite{Huang:2008}. 
Rather, nodes communicate as soon as they have the opportunity to do it. These occasional contacts are employed to share and disseminate information, route messages towards some destination, etc. Delay tolerance is thus a main characteristics of these networks, and in general the temporary contacts among nodes must be exploited to disseminate contents, due to the uncertainty of future communications.
% In this sense, the use of epidemic protocols is an interesting option, since it allows to rapidly disseminate information by exploiting the temporary contacts among nodes.

% In \cite{chaintreau}, for instance, opportunistic networks are modeled as small-worlds, where contacts between nodes may occur uniformly at random and with fixed probability, the number of nodes is considered as large (tending to infinity), and in general no mobility assumptions are made in the model. In \cite{phanse}, instead, it is argued that temporal connection models are better suited than spatial mobility models.

% Opportunistic mobile networks differ from classic mobile ad-hoc networks. The latter ones are usually managed to build a stable overlay which could be periodically re-configured to react to important topological updates in the spatial distribution of mobile nodes. Conversely, in opportunistic networks the evolving nature of the network topology is seen as a feature to exploit, rather than an issue to cope with 
% \cite{chaintreau,aict,phanse}. 

An open research topic is concerned with the topology of an opportunistic network. Due to its evolving nature, several works simply define the topology of opportunistic networks as unpredictable, others approximate them as small worlds, others argue that temporal connection models are better suited than spatial mobility models \cite{boldrini,chaintreau,phanse}.
As a matter of fact, at a given instant an opportunistic network appears as a classic mobile ad-hoc network (MANET), composed of a set of links which are constrained by the geographical location of mobile users and the limitations imposed by the signal strength of the employed wireless technology \cite{mobiopp}. 
However, while MANETs consider links as connections active at a given instant, opportunistic networks have a coarse grained time model.
Hence, the concept of link in an opportunistic network should reflect the fact that temporal constraints are relaxed.
This suggests to model opportunistic networks as evolving graphs where links characterize the interactions of a node during a (non-instantaneous) time interval $\Delta$.
Thus, the set of links of a node at time $t$ is not simply composed of the contacts active at the instant $t$, but rather, it represents the aggregate of the contacts arose during $[t-\Delta, t]$.
Then, different specific types of ``link'' are possible.
For instance, one might consider a link between two nodes as a single contact occurred during $\Delta$, or when a quasi-continuous communication is guaranteed between the two nodes (during $\Delta$).

A clear consequence of such a scenario is that links among
nodes would not depend only on spatial constraints, but also
on the mobility of the node and on the length of $\Delta$. In fact,
certain mobile users may be more active than others; hence,
while at a given instant the number of contacts of a highly-
mobile node may be similar to those of other nodes, during
the interval $\Delta$ such node will collect more contacts, resulting
in a higher number of links.

% A clear consequence of such a scenario is that links among nodes would not depend only on spatial constraints, but also on the mobility of the node and on the length of $\Delta$. In fact, certain mobile users may be more active than others; hence, while at a given instant the number of contacts of a highly-mobile node may be similar to those of other nodes, during the interval $\Delta$ such node will collect more contacts, resulting in a higher number of links.

In this scenario, it might be interesting to understand
whether the network can be shaped to achieve a desired
topology, based on some link management scheme, locally
executed by mobile nodes. The goal is clear: knowing the
network topology corresponds to identifying its statistical
characteristics. Based on these properties, it is possible to
define local strategies (e.g. which and how many links to
maintain) to optimize the opportunistic dissemination
% It might be thus interesting to understand whether the network can be shaped to achieve a desired topology, based on some link management scheme, locally executed by mobile nodes.
% The goal is clear: knowing the network topology corresponds to identifying its statistical characteristics. Based on these properties, it is possible to define local strategies to optimize the performances of the network in terms of communication responsiveness and reliability 
\cite{gridpeer}.

With the aim of identifying a strategy to shape opportunistic networks based on a given desired topology,
we define a method to control the number of links at nodes. In our tests we focus on a scale-free topology. The reasons for this choice are the following ones. First, it has been recognized that a scale-free network has several interesting characteristics for unstructured networks (e.g.~peer-to-peer), such as low diameter and high resilience to faults. Second, this topology is quite different to that typical of a network with spatial constraints, such as a MANET; the achievement of such a desired topology entails that several other opportunities are possible. 

A simulation environment was built to test the model and the local algorithm to be executed at mobile nodes.
Experiments are based on real traces taken from \cite{nus-contact-2006-08-01}, which report the typical everyday behavior of more than 20000 students of the University of Singapore, equipped with a mobile terminal. These traces represent a perfect case study for an opportunistic network.
Results on these traces demonstrate that, by tuning the length of the time interval $\Delta$, we can shape the opportunistic network based on some desired topology, e.g.~scale-free.

The main contributions of this work are the following ones. 
We characterize opportunistic networks as evolving graphs where links correspond to active contacts during a time interval. We present a simple scheme to shape the network based on some desired topology. We specifically cope with scale-free topologies, but the approach can be easily generalized to other network topologies. Finally, we experiment with real data-traces, which record the typical behavior of a large number of users during their everyday life \cite{nus-contact-2006-08-01}.

The remainder of the paper is organized as follows. Section \ref{sec:alg} presents the general approach. Section \ref{sec:exp} outlines the experimental scenario, while results are reported in Section \ref{sec:res}.
Finally, Section \ref{sec:conc} provides some concluding remarks.

%%%%%%%%%%%%%%%%%%%%
%% presento l'idea 
%%%%%%%%%%%%%%%%%%%%

\section{Characterization of Opportunistic Networks}
\label{sec:alg}

We consider an opportunistic network as an evolving graph where links among nodes correspond to the aggregate of contacts that nodes have during a given time interval $\Delta$. Thus, at a given instant a node is in direct contact with a subset of those nodes linked to it during the $\Delta$ time window.
% Definition of instantaneous network topology ???
We then present a simple scheme that lets nodes to maintain a pre-defined number of neighbours. This enables to set a ``desired topology''. We employ this mechanism in the attempt to give to the network a scale-free desired topology.

\subsection{Towards a Model for Opportunistic Networks}

Opportunistic networks are characterized by nodes that frequently move and with intermittent contacts \cite{Huang:2008}. An opportunistic network may assume different forms and it may be employed as the underlying structure for several mobile application scenarios.
Applications built on top of opportunistic networks are different to those to be executed in classic (wireless) networks.
Examples may be found in literature, ranging from communication approaches based on ad-hoc networks to be deployed on rural and developing regions, tracking systems for wild animals, to those (delay tolerant) applications built over highly dynamic nets in urban areas (e.g.~mobile and vehicular ad-hoc networks) \cite{conti,Huang:2008,Lindgren:2009,qiang}.
In general, no simultaneous multi-hop path can be guaranteed \cite{conti}. 
Communication protocols allow nodes to manage intermittent and unstable contacts, cache and relay messages as soon as there is the opportunity to do it. From an implementation point of view, there is a ``bundle''-layer, in between the classic transport and the application layers, that implements such store, carry and forward strategy.
% As to the communication protocols among nodes, instead, the topology of the network is very dynamic; hence, building stable overlays (as thought for MANETs) to organize the communication among nodes may be a costly and unsuccessful operation.

A critical point is the definition of ``link'' between nodes. This must certainly be based on contacts between two nodes and might depend on the type of interactions required by the applications run on top of the network itself \cite{Lindgren:2009}.
For instance, a single contact is sufficient to exchange some content of common interest between two users, e.g.~variations on lecture schedules among students, news on train/airport tables, and in general, content distribution through some mobile publish-subscribe like service. In this case, we may state that a link exists between two nodes during a time interval $\Delta$, if a contact arises between the two nodes within that time interval.

% Therefore, the routing of data can be made at different time instants, after changes of the instantaneous network topology that offer novel nodes to communicate with.

We hence consider the network as an evolving graph whose links of a node $i$ at time $t$ are the aggregate of active contacts of $i$ during the time interval $[t-\Delta, t]$ \cite{Monteiro:2006}.
We assume that only messages received in a $\Delta$ time window can be relayed to a neighbour though an active link. This constraint reflects the fact that there is a limit on the temporal validity of messages and also to consider the limited computational capabilities of mobile nodes which cannot hold messages forever.
% We now present a simple scheme that is based on a local control of the degree of a mobile node in an opportunistic network.

The implementation of a link, as thought in this context, may be realized at the bundle and application layers, rather than as an open communication flow between two nodes (as usually thought in MANETs or P2P architectures). In practice, a link from a node $i$ to another node $j$ means that $i$ stores on its memory the information related to $j$, such as $j$'s id and its profile, together with the contents $j$ is trying to disseminate, or the type of contents $j$ is trying to retrieve (depending on the service running on top of the opportunistic network).
This approach of modeling a link reflects the fact that some content disseminated from $i$ to $j$ may be delivered afterwards to another node $k$, if $j$ has the opportunity to interact with it in a time interval $\Delta$.
Due to constraints of mobile devices, each node would maintain a limited number of nodes as neighbours, and a node might decide to replace a neighbour with another.
Hence, it may happen that as the network evolves, a node $i$ has an entry for $j$ in its ``neighbour table'', while $j$ does not have any entry for $i$, i.e.~links are directed.

\subsection{Shaping Opportunistic Networks}

We define the desired topology of an opportunistic network by specifying the probability distribution of the degree (i.e.~number of links) that a node may have. Through this choice, it is possible to give a statistical characterization of the network, which permits to automatically estimate important metrics, such as the net diameter (i.e.~the max of the shortest paths among nodes in a net), the average number of neighbours at a given distance from a specific node, and so on \cite{gridpeer}.

The algorithm for the link management is very simple and it is as follows. 
At the beginning of its interactions, each node randomly selects a desired degree, locally computed through the degree probability distribution associated to the desired topology. Note that in order to randomly selecting a degree, the node might need an estimation of the network size. Schemes exist that do it \cite{Kostoulas:2007,lemerrer}. Moreover, when the network size is very large, just an approximate value is sufficient; it is required that, in case of a significant variation, nodes can detect it, and in this case they might change their desired degree.
During the evolution of the network, based on the active contacts a node has, the node stores entries related to the nodes it encounters (i.e.~it creates links), till reaching its desired degree. 

In general, we assume that links can be established only when the number of active contacts (within a time interval $\Delta$) surpasses a predefined threshold $c_{min}$. This parameter can be tuned depending on the type of service to be executed on top of the opportunistic network. For instance, a single contact ($c_{min}=1$) may be employed for simple dissemination services where contents must be broadcast through the net. Higher values might be set for more sophisticated services, e.g.~queries to be distributed that need answers. In this case, the communication would require multiple contacts.

The link holds for a limited time, and it is removed after a time $\Delta$ of no contacts.
When a novel contact arises with a non-neighbour node (say $j$), and if the node has already reached its desired degree, then the novel node may become a neighbour with a certain probability $\omega$. In this case, a random entry (e.g.~one among those with the lowest number of active contacts during the $\Delta$ time window) is replaced with $j$.

\subsection{A Study with Scale-free Nets}
\label{sec:scale}

In the next section we evaluate whether an opportunistic network, modeled as described above, can assume a scale-free topology. A scale-free network 
possesses the distinctive feature of having nodes with a degree distribution that can be well approximated by a power law function. Hence, the majority of nodes have a relatively low number of neighbors, while a non-negligible percentage of nodes (``hubs'') exists with higher degrees \cite{simutools}. 
The peculiarity of these networks is that they possess a very small diameter, thus allowing to propagate information in a low number of hops. They are quite robust to node faults (departures), something usual in a wireless network. 
On the other hand, the presence of hubs might represent a drawback in the context of opportunistic networking, since it corresponds to an unbalanced load distribution.

Coupling scale-free and (opportunistic) mobile networks is unusual, due to the issue mentioned above and to the fact that in a MANET nodes connect to those which are directly reachable through the networking technology in use. Therefore, the instantaneous topology strongly depends on the geographical distribution of the nodes.
However, in this work links are considered as the aggregate of active contacts arising during a time interval $\Delta$. 
Hence, the role of hubs might be played by those nodes that have in time a number of contacts higher than others. An example of a possible hub, in a real opportunistic network, might be a ticket inspector in some public transportation system (equipped with a mobile terminal), a postman, or even a dedicated totem (i.e.~``information sprinkler'') placed to relay contents to mobile nodes in a square or in a mall \cite{Bulander:2005,Ratsimor:2003}. The idea of having dedicated nodes in points of interest, which act as hubs able to relay contents, would solve also the issue of the unfair load at hubs.

In any case, while we evaluate whether a scale-free overlay can be built over an opportunistic network, the same machinery can be employed also for other topologies. It suffices to change the distribution to compute the desired degree.

%%%%%%%%%%%%%%%%%%%%
%% Evaluation
%%%%%%%%%%%%%%%%%%%%
\begin{figure}[t]
   \centering
   \includegraphics[angle=-90,width=.8\linewidth]{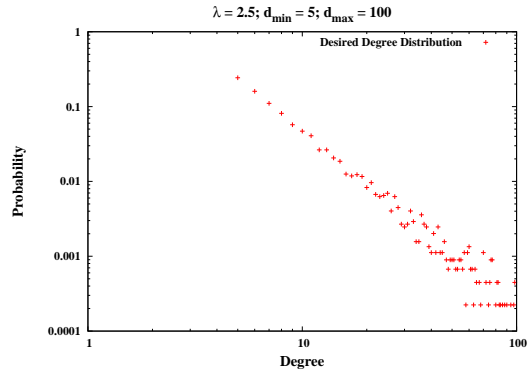}
   \caption{Example of the desired degree distribution of a network generated during the tests. $\lambda=2.5, d_{min}=5, d_{max}=100$}
   \label{fig:dd}
\vspace{-0.5cm}
\end{figure}

\section{Experimental Evaluation}
\label{sec:exp}

A discrete event simulator was built to mimic the algorithm executed at mobile nodes. Nodes' movements and contacts were obtained from real data traces \cite{nus-contact-2006-08-01}. In what follows, we describe these traces and explain the metrics of interest, together with the methodology to collect the results, which are provided in the next section.

\subsection{Data Traces}

We employed real data traces available from \cite{nus-contact-2006-08-01}. These are the patterns among 22341 students, inferred from their behavior during class schedules for the Spring semester of 2006 in National University of Singapore. 
The National University of Singapore is composed of colleges and departments. As reported in the description of the traces, all lessons were conducted on the main campus, spanning an area of 146 hectares \cite{nus-contact-2006-08-01}.
In essence, these traces represent a wide, real scenario of typical students' everyday life. It is hence an ideal use case to test algorithms on urban opportunistic networks. 

User contacts were traced in time intervals. For instance, students meeting during a lecture where considered to be in contact during a single time interval,
regardless of its duration (which is abundantly longer than time usually needed to exchange information between two mobile nodes). Hence, the time interval $\Delta$ exploited to manage links in the network is expressed as a multiple of these intervals.

% The rule is simple - 
% two students are in contact with each other if and only if they are
% in the same venue at the same time. In other words, we assume
% that as long as two students are in the same classroom, they are
% within Bluetooth range of each other. This assumption has been
% validated inside large classrooms on our campus. We also assume
% that two students who are in different classrooms are out of range
% of each other, even if one classroom is just next door to the other.
% We further assume that contacts take place only during business
% hours, and ignore that fact that students hang around campus for
% various activities after hours. We note that the last two assumptions
% are conservative - the number of contacts we obtained is a lower
% bound of the actual contacts that take place on campus.

\subsection{Metrics and Methodology}

The desired topology was generated through a probability distribution for nodes' desired degree $d$, following a power law function $\sim d^{\lambda}$, with $\lambda \in [-3,-2]$.
We considered a minimum and a maximum value of the desired degrees that a node might want to have, $d_{min}$, $d_{max}$, respectively. We varied these parameters.
Figure \ref{fig:dd} shows an example of the distribution of the desired degree of network nodes, generated during our tests. The log-log chart shows a linear curve, hence confirming that such distribution follows a power law distribution.

During our tests we varied the time interval size $\Delta$. 
% In particular, $\Delta$ was set as a multiple of intervals employed in the traces.
We varied also the minimum number of active contacts $c_{min}$ required so that to two nodes may become neighbours, and the value of $\omega$ that controls the probability that a node replaces a link with another, once its degree is equal to its desired degree.

%%%%%%%%%%%%%%%%%%%%
%% Results
%%%%%%%%%%%%%%%%%%%%
\section{Results}
\label{sec:res}

\begin{figure*}[t]
   \centering
   \includegraphics[angle=-90,width=.4\textwidth]{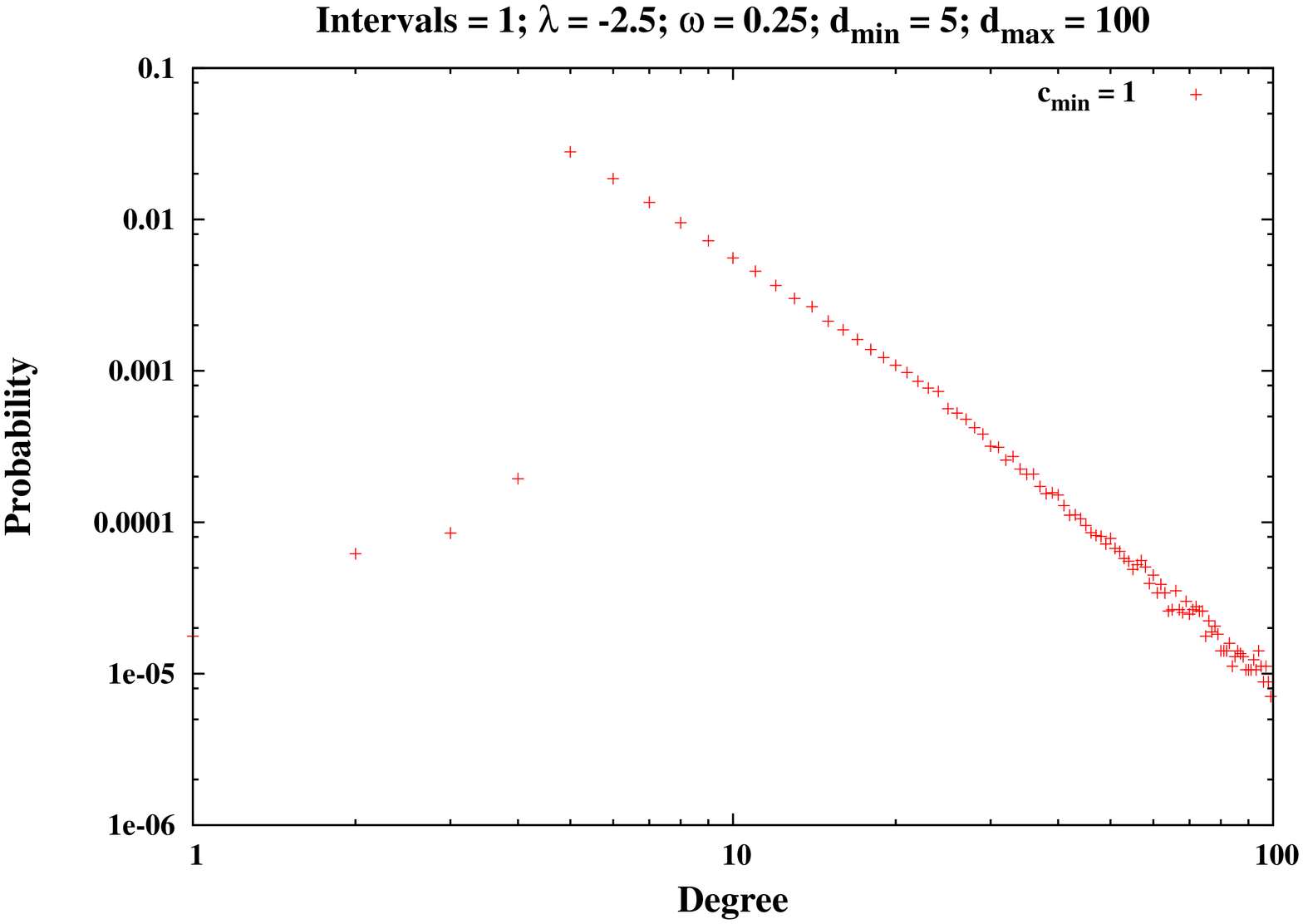}
   \includegraphics[angle=-90,width=.4\textwidth]{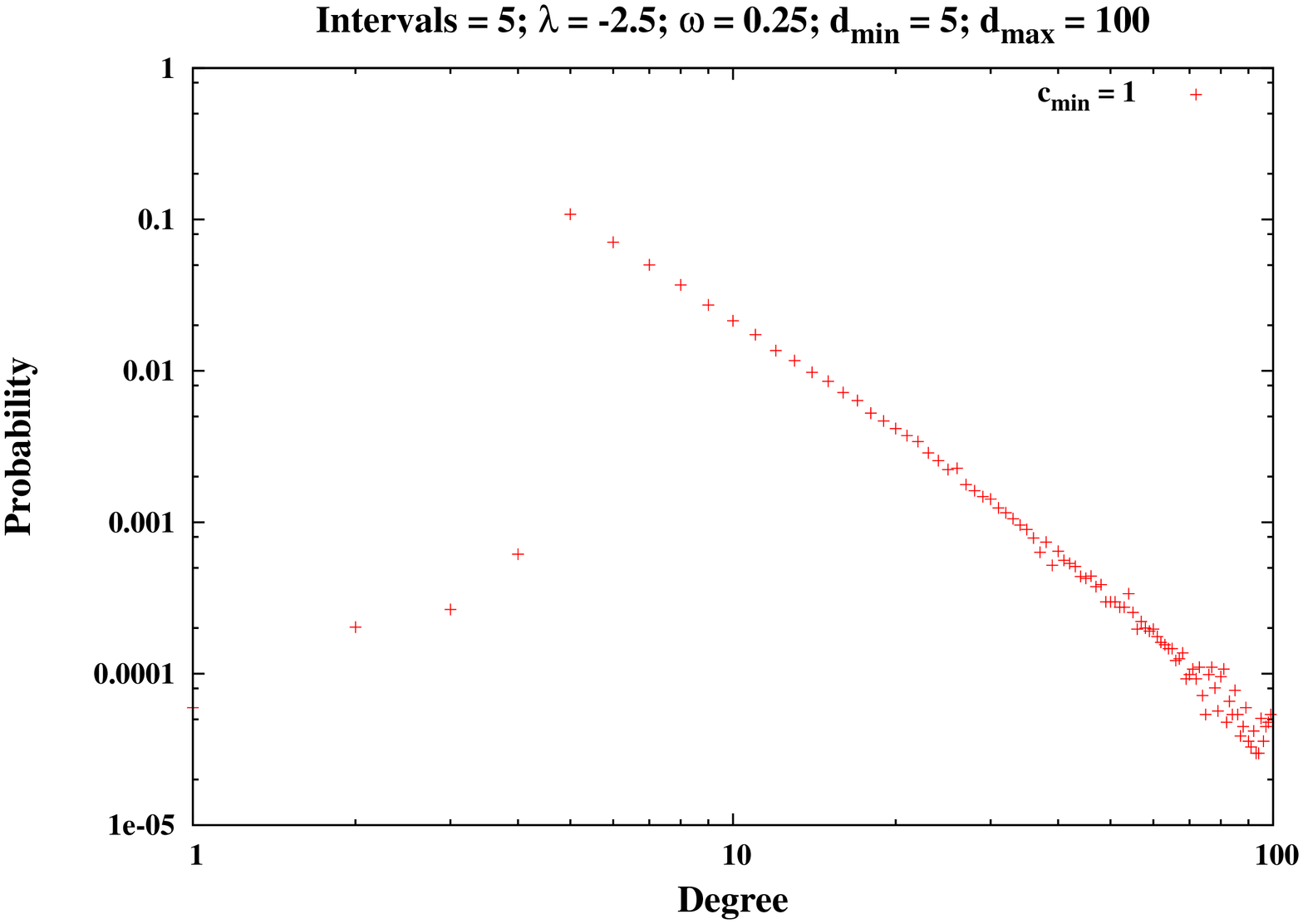}
   \caption{Degree Probability; time interval duration varied; $d_{min}=5$}
   \label{fig:res_min_5}
\vspace{-0.5cm}
\end{figure*}

\begin{figure*}[t]
   \centering
   \includegraphics[angle=-90,width=.4\textwidth]{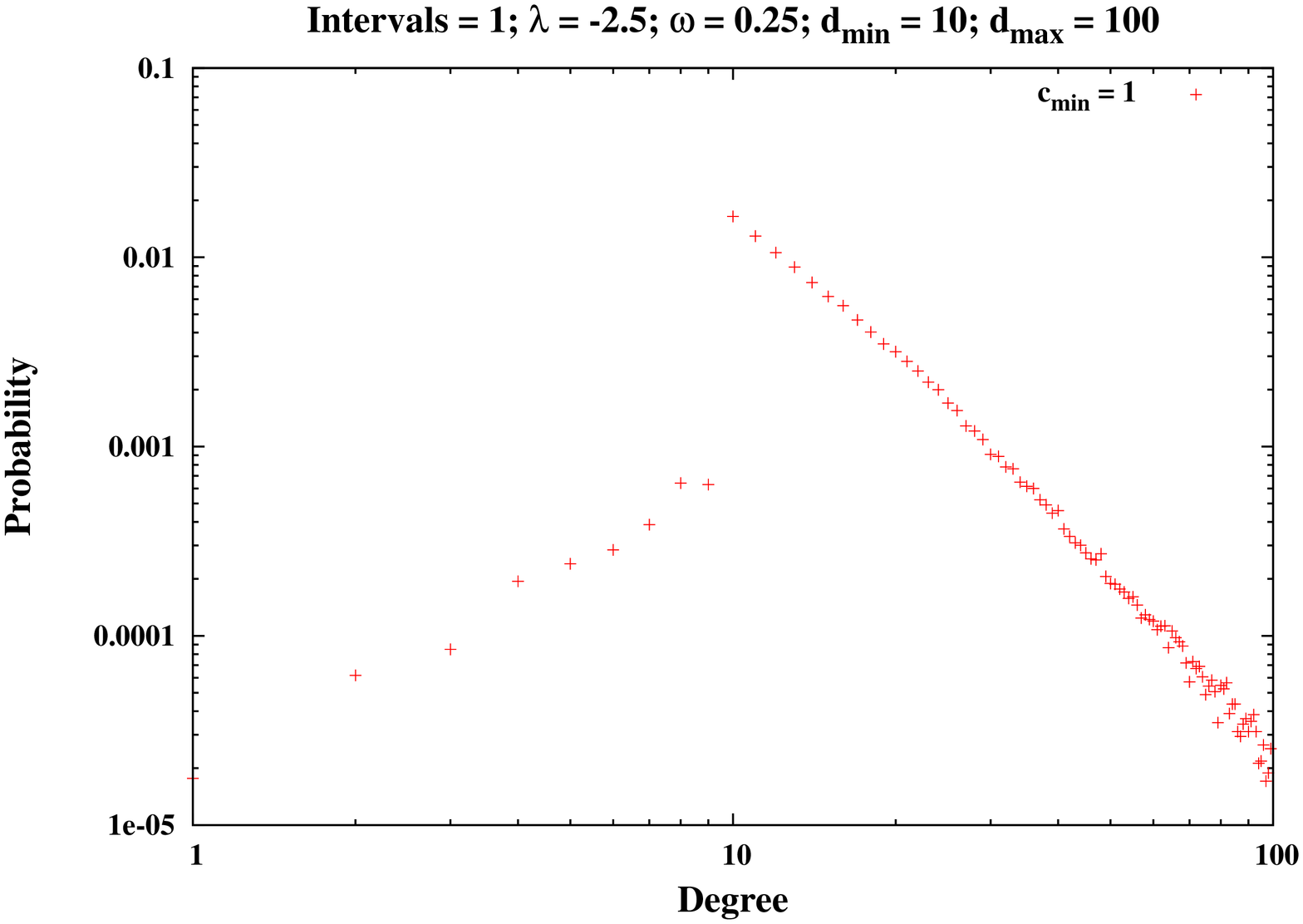}
   \includegraphics[angle=-90,width=.4\textwidth]{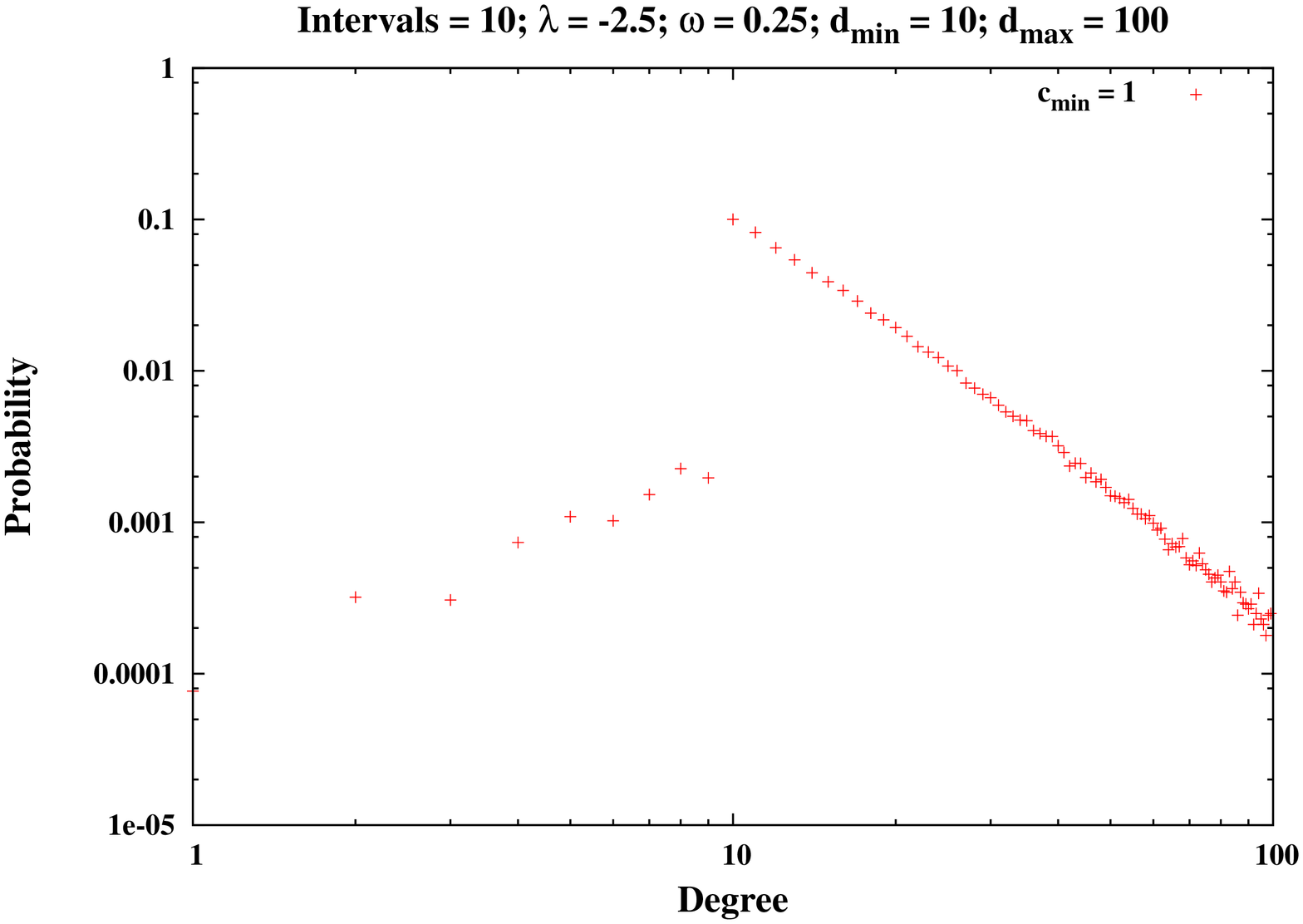}
   \caption{Degree Probability; time interval duration varied; $d_{min}=10$}
   \label{fig:res_min_10}
\vspace{-0.5cm}
\end{figure*}

%%%%%%%%%%%%%%%%%%%%%%%%%%
\begin{figure*}[t]
   \centering
   \includegraphics[angle=-90,width=.4\textwidth]{5_5_100_-2.5_0.25.ps}
   \includegraphics[angle=-90,width=.4\textwidth]{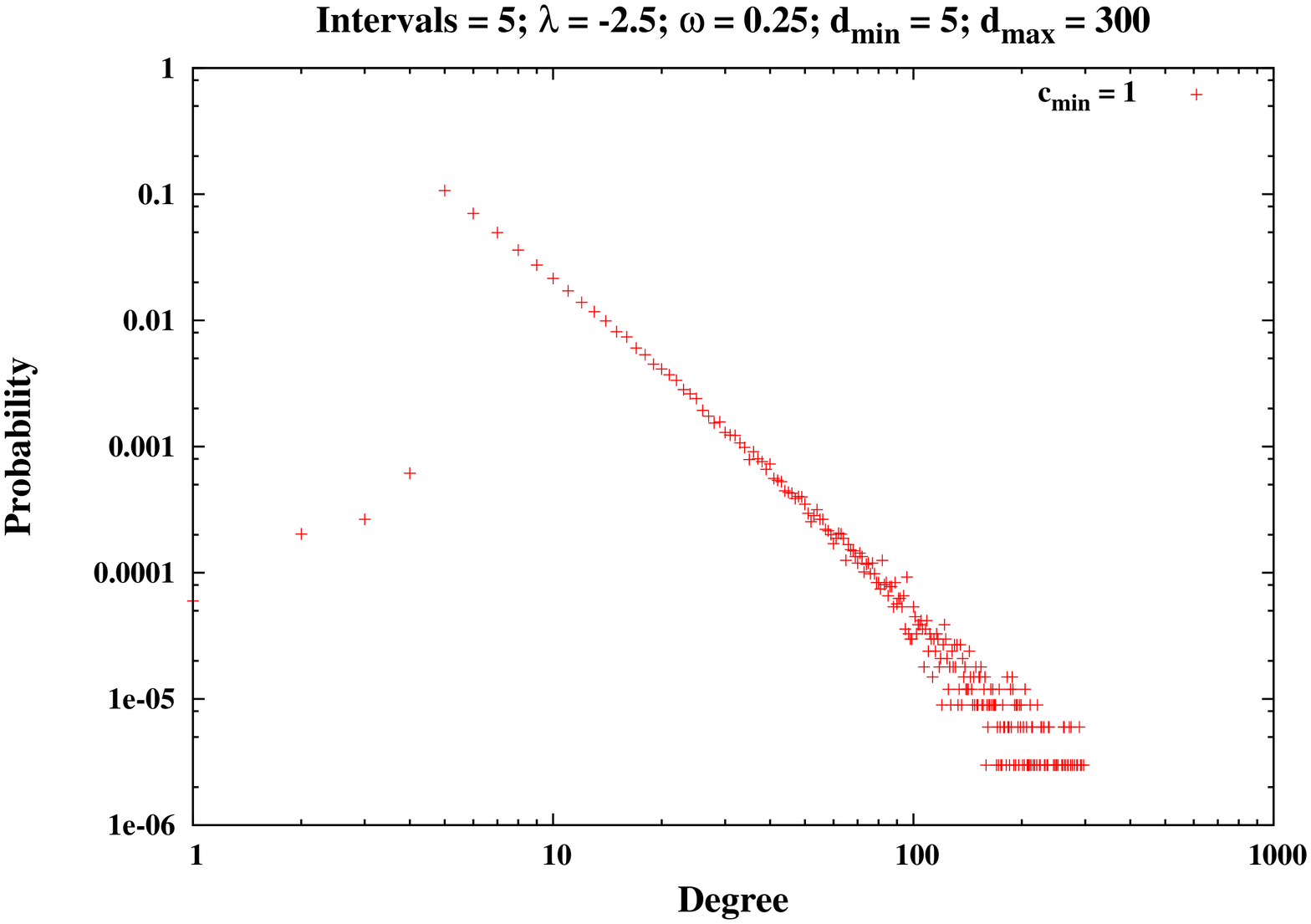}
   \includegraphics[angle=-90,width=.4\textwidth]{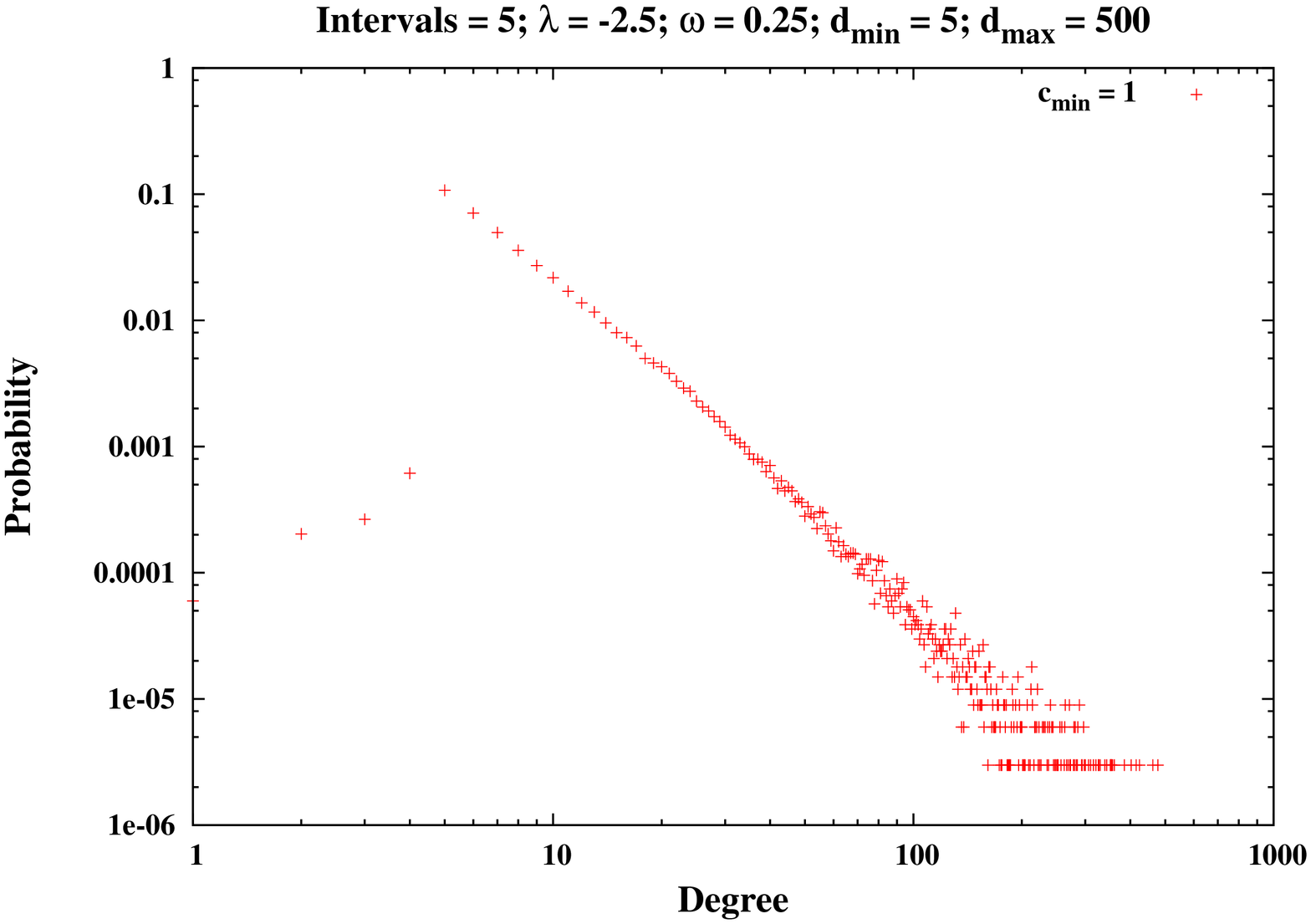}
   \includegraphics[angle=-90,width=.4\textwidth]{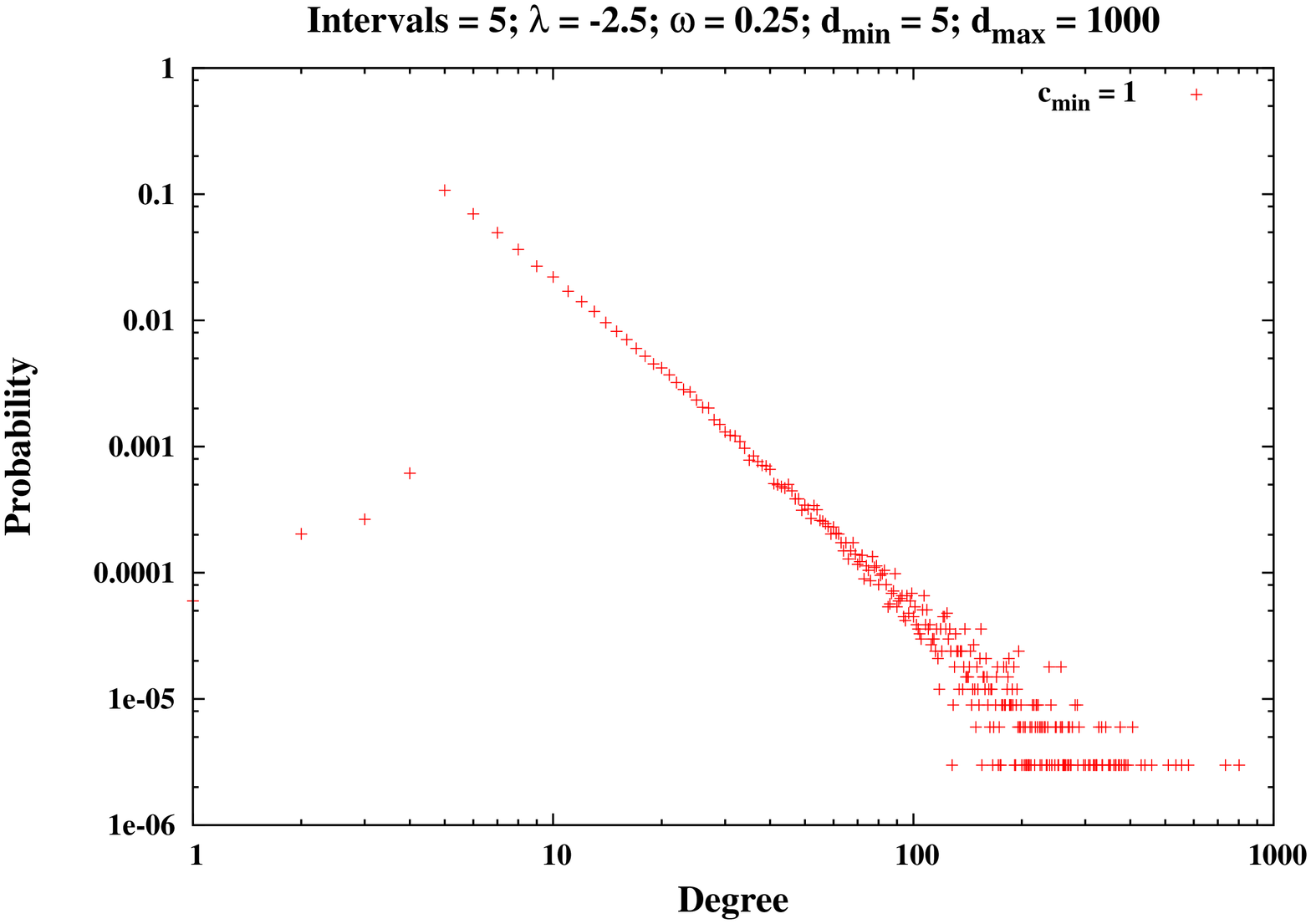}
   \caption{Degree Probability; $d_{max}$ varied; $d_{min}=5$}
   \label{fig:res_d_max_variato}
\vspace{-0.5cm}
\end{figure*}

For the sake of conciseness, we present only results related to a setting with 
$\lambda = -2.5$, 
$\omega = 0.25$,
$c_{min} = 1$,
while varying the values $d_{min} \in \{5, 10\}$, $\Delta \in \{1, 5, 10\}$, $d_{max} \in \{100, 300, 500, 1000\}$.
Similar outcomes were obtained for other values of these parameters (not shown in this paper).

It goes without saying that, depending on the service to be executed on top of the opportunistic network, and most of all, on the contents to be disseminated, a degree value such as that of $d_{max} \sim 1000$ might be too high for a mobile node. Other scenarios are probably more realistic. It is however worth mentioning that: i) the number of such hubs would be very low and in certain application scenarios such a role might be played by those ``information sprinklers''; ii) this setting is a test to see whether such a threshold might be reached using a real trace such as that employed in these tests.

Figure \ref{fig:res_min_5} reports the results when $d_{min}=5$, $d_{max} = 100$, varying the duration of $\Delta$, (i.e.~$\Delta = 1, 5$ trace intervals). Each chart refers to a particular value of $\Delta$, and reports the distribution of nodes' degree obtained when the proposed algorithm is executed. The probability that a node has a certain degree is reported in a log-log scale. 
In this case, a linear trend may be appreciated, meaning that the distribution follows a power law function. This confirms that the approach is able to configure the topology as a scale-free. 
Moreover, an higher $\Delta$ (rightmost chart) guarantees a higher probability that nodes reach their desired degree. 
Only some few non-zero probabilities are obtained for some degree values lower than $d_{min}$. This means that there are certain nodes that have a few number of contacts with other nodes, whatever the duration of $\Delta$. 
A similar result is reported in Figure \ref{fig:res_min_10}, where $d_{min} = 10$ and $\Delta = 1, 10$ trace intervals.

% Results show that with an adequate length of the time interval $\Delta$, the degree distribution has a linear trend in the log-log scale, meaning that it follows a power law function (and hence, a scale-free network is obtained). This is valid for each $c_{min}$ value, except for the case $c_{min}=1$ with degrees lower than $d_{min}$. In practice, there are nodes that have a degree which is lower than its desired degree (which is greater or equal than $d_{min}$); for the unwanted values lower than $d_{min}$, the probability increases till reaching the $d_{min}$ threshold; then the probability decreases in a linear fashion, as expected.

Figure \ref{fig:res_d_max_variato} reports the degree probability obtained when $\Delta = 5$ trace intervals, $d_{min}=5$, while varying $d_{max}$.
Again, it is confirmed that a scale free topology inside an opportunistic network is obtained in all cases, since there is a linear distribution of degree probabilities (in log-log scale). When $d_{max} = 1000$, the probability that a node has a high degree is very low. This is however due to the fact that only few nodes might have the opportunity to meet a number of $\sim1000$ nodes during a $\Delta$ time window. Moreover, such values are inappropriate as number of links a mobile node might have in a real opportunistic network; such simulation setting serves only to demonstrate that the framework scales up to those numbers.
The reader may also notice the presence of non-zero degree probabilities lower than $d_{min}$, which remain, despite the settings of the algorithm executed (i.e.~value of $\Delta$). The presence of these values comes from the fact that the data traces exploited to do the simulations were the same in all the considered scenarios.

% As shown in the results, degree probabilities do not reach the value $d_{max}$ set in these tests. This is simply due to the fact that the employed data traces do not contain such a high number of contacts. Hence, during the tests those nodes with a desired degree equal to $d_{max}$ alway create novel links without deleting any contact, during a single time interval. In any case, the obtained results confirm that nodes can maintain a number of links which follow a power law distribution, thus creating a scale free topology inside an opportunistic network.

%%%%%%%%%%%%%%%%%%%%
%% CONCLUSIONS
%%%%%%%%%%%%%%%%%%%%

\section{Conclusions}
\label{sec:conc}

In this work we have presented a discussion on how opportunistic networks might be modeled. Due to the evolving nature of the network and its delay tolerance, links among nodes should not be considered as contacts which are active simultaneously. Rather, the aggregate of contacts arising during a time interval should be preferred. 
To create a link, the
duration of the contact might be sufficient for a data exchange.
Such amount of time might depend on the application to
be run on top of the network. In particular, a link in the
evolving graph represents the fact that a node maintains on
its ``neighbour table'' some application-related contents on the
behalf of its neighbour, so that such contents can be relayed
as soon as there is the opportunity to do it.
% Then, a link corresponds to some contact between two nodes, whose duration is sufficient for a data exchange. Such amount of time might depend on the application to be run on top of the network. 
% In particular, a link in the evolving graph represents the fact that a node maintains on its ``neighbour table'' some application-related contents on the behalf of its neighbour, so that such contents can be relayed as soon as there is the opportunity to do it.

% Taking into account such an approach, 
We have presented a simple algorithm that allows each node to set and manage its own degree, in order to shape the net based on a desired topology. This influences the way contents can be disseminated through the network.
The scheme has been employed on real data traces, and outcomes confirm that a desired topology (in this case a scale-free) can be obtained.

To the best of our knowledge, this is the first attempt to model opportunistic networks taking into consideration contacts active in a given time interval, instead of instantaneous contacts. In future works, the scheme will be employed on other data traces and with different desired topologies.

%%%%%%%%%%%%%%%%%%%%
%% BIBLIOGRAFIA
%%%%%%%%%%%%%%%%%%%%

\small{

}

% \balancecolumns

\end{document}